\documentclass[eqsecnum,twocolumn,aps,epsf]{revtex4}
\usepackage{graphicx}
\usepackage{color}
\begin{document}
\draft

\title{Magnetic-field effects on the in-plane electrical resistivity in the single-crystal La$_{2-x}$Ba$_x$CuO$_4$ and La$_{1.6-x}$Nd$_{0.4}$Sr$_x$CuO$_4$ around $x=1/8$:\\Relating to the field-induced stripe order}

\author{T. Adachi, N. Kitajima, T. Manabe, Y. Koike}

\address{Department of Applied Physics, Graduate School of Engineering, Tohoku University, \\Aoba-yama 6-6-05, Aoba-ku, Sendai 980-8579, Japan}

\author{K. Kudo, T. Sasaki, N. Kobayashi}

\address{Institute for Materials Research, Tohoku University, Katahira 2-1-1, Aoba-ku, Sendai 980-8577, Japan}

\date{\today}

\begin{abstract}

Temperature dependence of the in-plane electrical resistivity, $\rho_{\rm ab}$, in various magnetic fields has been measured in the single-crystal La$_{2-x}$Ba$_x$CuO$_4$ with $x=0.08$, 0.10, 0.11 and La$_{1.6-x}$Nd$_{0.4}$Sr$_x$CuO$_4$ with $x=0.12$. 
It has been found that the superconducting  transition curve shows a so-called fan-shape broadening in magnetic fields for $x=0.08$, while it shifts toward the low-temperature side in parallel with increasing field for $x=0.11$ and 0.12 where the charge-spin stripe order is formed at low temperatures. 
As for $x=0.10$, the broadening is observed in low fields and it changes to the parallel shift in high fields above 9 T. 
Moreover, the normal-state value of $\rho_{\rm ab}$ at low temperatures markedly increases with increasing field up to 15 T. 
It is possible that these pronounced features of $x=0.10$ are understood in terms of the magnetic-field-induced stabilization of the stripe order suggested from the neutron-scattering measurements in the La-214 system. 
The $\rho_{\rm ab}$ in the normal state at low temperatures has been found to be proportional to ln(1/$T$) for $x=0.10$, 0.11 and 0.12. 
The ln(1/$T$) dependence of $\rho_{\rm ab}$ is robust even in the stripe-ordered state.

\end{abstract}
\vspace*{2em}
\maketitle
\newpage

%*****************************************************************************************
\section{Introduction}~\label{intro}
%*****************************************************************************************
Magnetic fields can tune the electronic state of strongly-correlated electron systems. 
It is well-known that magnetic fields strongly affect the superconducting properties as well as the normal-state ones. 
For conventional superconductors, the superconducting transition curve in magnetic fields shifts to the low-temperature side in parallel with increasing field, which is attributed to the small superconducting fluctuation originating from the large superconducting coherence length. (This is called a parallel shift.)
For the high-$T_{\rm c}$ cuprates, on the other hand, the superconducting transition curve shows a so-called fan-shape broadening in the underdoped regime,~\cite{kitazawa,suzuki} which is attributed to the large superconducting fluctuation originating from the small superconducting coherence length and the quasi-two-dimensional superconductivity. 

As for magnetic-field effects on the normal-state properties, some interesting behaviors of the normal-state electrical resistivity at low temperatures below the superconducting transition temperature $T_{\rm c}$ have been found through the destruction of the superconductivity by the application of magnetic field. 
In the underdoped La$_{2-x}$Sr$_x$CuO$_4$ (LSCO), for example, the in-plane resistivity, $\rho_{\rm ab}$, in the normal state exhibits an insulating behavior at low temperatures, diverging in proportion to ln(1/$T$), though the origin of the ln(1/$T$) dependence has not been clarified.~\cite{ando,boebinger} 
The ln(1/$T$) dependence has also been observed in the underdoped Bi$_2$Sr$_{2-x}$La$_x$CuO$_{6+\delta}$ (BSLCO) below $p$ (the hole concentration per Cu) $\sim 0.12$,~\cite{ono} suggesting that the ln(1/$T$) dependence may be a common feature of $\rho_{\rm ab}$ in the normal state of the underdoped high-$T_{\rm c}$ cuprates at low temperatures. 

Recently, magnetic-field effects on the charge-spin stripe order~\cite{nature,prb} have also attracted great interest. 
Elastic neutron scattering measurements in magnetic fields for LSCO with $x=0.10$ (Ref.~\cite{lake}) under the orthorhombic mid-temperature (OMT) structure (space group: {\it Bmab}) have revealed that the intensity of the incommensurate magnetic peaks around ($\pi$, $\pi$) in the reciprocal lattice space increases with increasing field parallel to the c-axis, suggesting the stabilization of the magnetic order. 
For LSCO with $x=0.12$, on the other hand, the enhancement of the incommensurate magnetic peaks is observable but small.~\cite{katano} 
The magnetic order is considered to be almost stabilized even in zero field for $x\sim1/8$,~\cite{suzuki-san,kimura} so that the development of the magnetic order by the application of magnetic field is slight. 
The enhancement of the incommensurate magnetic peaks has also been observed for the excess-oxygen doped La$_2$CuO$_{4+\delta}$ (LCO) with the stage-4 and state-6 structures.~\cite{khay,khay2} 
For La$_{1.6-x}$Nd$_{0.4}$Sr$_x$CuO$_4$ (LNSCO) with $x=0.15$ where the charge-spin stripe order is stabilized under the tetragonal low-temperature (TLT) structure (space group: {\it P}4$_2$/{\it ncm}), on the contrary, field effects on neither the charge nor magnetic peaks associated with the charge-spin stripe order have been observed up to 7 T.~\cite{wakimoto} 
These results suggest that a sort of spin stripe order in the OMT phase of LSCO and in LCO is stabilized by the application of magnetic field, while the charge-spin stripe order in the TLT phase of LNSCO is almost never affected. 

\begin{figure}[tbp]
\begin{center}
\includegraphics[width=1.0\linewidth]{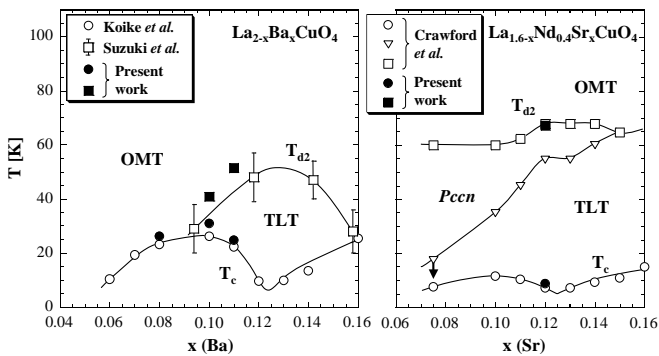}
\end{center}
\caption{(left) Phase diagram of La$_{2-x}$Ba$_x$CuO$_4$. Open circles represent $T_{\rm c}$, defined as the mid-point temperature in the resistive superconducting transition.~\cite{koike} Open squares represent the structural phase transition temperature between the OMT and TLT phases, $T_{\rm d2}$.~\cite{suzuki-fujita} (right) Phase diagram of La$_{1.6-x}$Nd$_{0.4}$Sr$_x$CuO$_4$.~\cite{crawford} Open circles represent $T_{\rm c}$. Open squares represent the structural phase transition temperature between the OMT and TLT/{\it Pccn} phases, $T_{\rm d2}$. Open triangles represent the temperature between the {\it Pccn} and TLT phases. Closed circles and squares in the both diagrams represent the present $T_{\rm c}$ and $T_{\rm d2}$ estimated from the $\rho_{\rm ab}$ measurements, respectively.} 
\label{p-d} 
\end{figure}

In this paper, with the aim to clarify the relation between the superconducting and normal-state properties and the formation of the stripe order, we have performed $\rho_{\rm ab}$ measurements in magnetic fields up to 15 T for the single-crystal La$_{2-x}$Ba$_x$CuO$_4$ (LBCO) with $x=0.08$, 0.10, 0.11 and LNSCO with $x=0.12$. 
As shown in Fig. \ref{p-d}, it is noted that both LNSCO with $x=0.12$ and LBCO with $x=0.11$ are located in the regime where the superconductivity is suppressed in the neighborhood of $p=x=1/8$, while LBCO with $x=0.08$ is located outside this regime. 
LBCO with $x=0.10$ is just at the boundary between the inside and outside of this regime. 
In the elastic neutron scattering measurements of LBCO~\cite{goka} and LNSCO~\cite{nature,prb}, the incommensurate elastic charge and magnetic peaks associated with the stripe order have been observed for $x=0.10$ and 0.12 below the structural phase transition temperature between the OMT and TLT/{\it Pccn} phases, $T_{\rm d2}$, in zero field, but not for $x=0.08$. 
These mean that the static charge-spin stripe order is formed at low temperatures below $T_{\rm d2}$ for $x=0.10$, 0.11 and 0.12 even in zero field. 
Moreover, it has been found that the intensity of the elastic charge peaks is weaker in $x=0.10$ than in $x\sim1/8$ of LBCO,~\cite{goka,fujita} suggestive of a less-stabilized static charge order in $x=0.10$. 
For $x=0.08$, on the other hand, the stripe order is not stabilized even at low temperatures. 

%*****************************************************************************************
\section{Experiments}
%*****************************************************************************************
Single crystals of LBCO with $x=0.08$, 0.10, 0.11 and LNSCO with $x=0.12$ were grown by the traveling-solvent floating-zone method under flowing O$_2$ gas of 4 bar. 
The detailed procedures are described elsewhere.~\cite{lbco-prb} 
Ba and Sr contents of each crystal were analyzed by the inductively-coupled-plasma (ICP) measurements. 
The $\rho_{\rm ab}$ was measured by the standard dc four-probe method on field cooling in magnetic fields parallel to the c-axis up to 15 T. 

%*****************************************************************************************
\section{Results}
%*****************************************************************************************
\begin{figure*}[tbp]
\begin{center}
\includegraphics[width=1.0\linewidth]{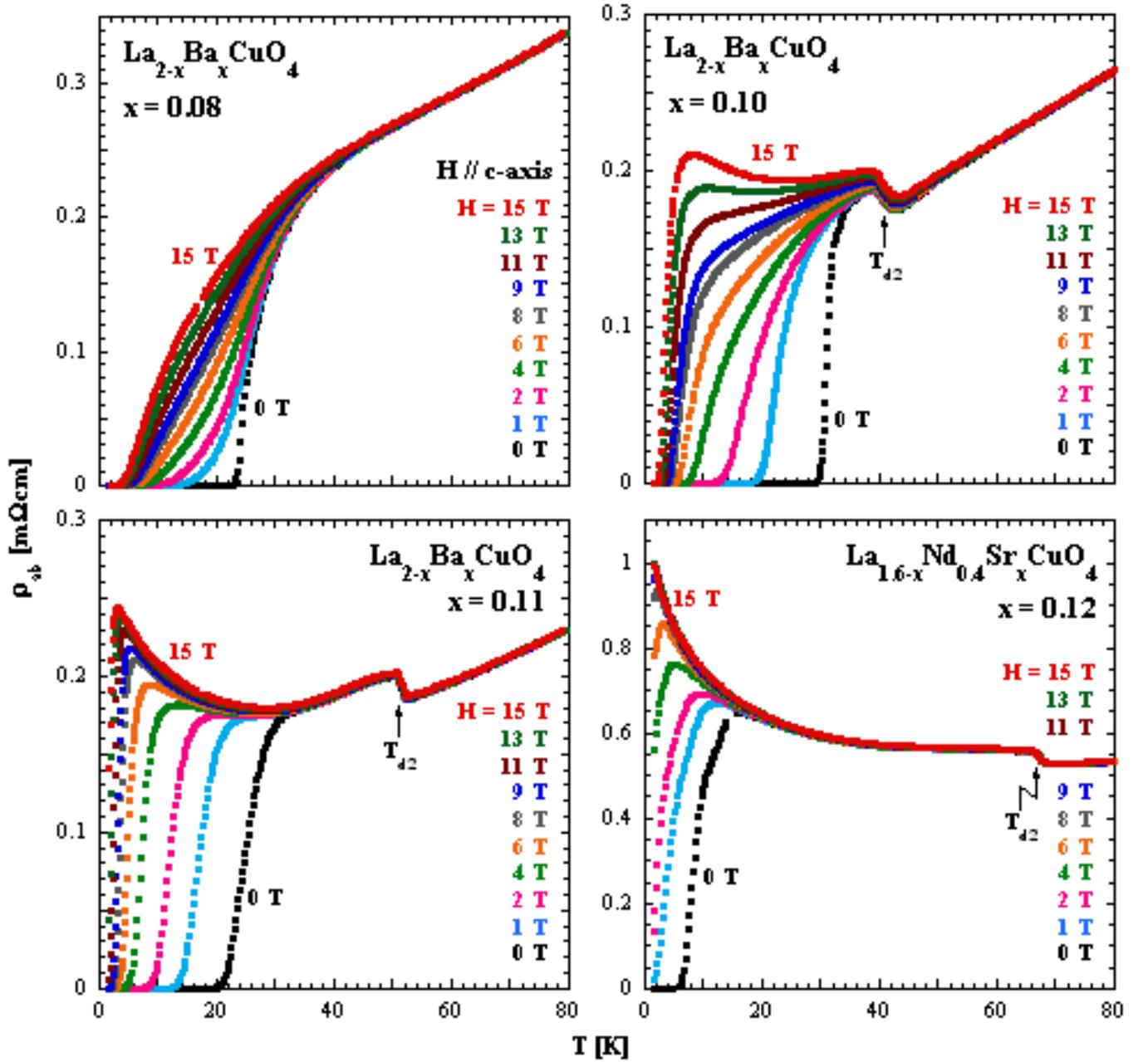}
\end{center}
\caption{(color online) Temperature dependence of the in-plane electrical resistivity, $\rho_{\rm ab}$, in various magnetic fields parallel to the c-axis for La$_{2-x}$Ba$_x$CuO$_4$ with $x=0.08$, 0.10, 0.11 and La$_{1.6-x}$Nd$_{0.4}$Sr$_x$CuO$_4$ with $x=0.12$. The temperature where a jump of $\rho_{\rm ab}$ occurs is in correspondence to the structural phase transition temperature between the OMT and TLT/{\it Pccn} phases, $T_{\rm d2}$.}  
\label{r-t} 
\end{figure*}

Figure \ref{r-t} displays the temperature dependence of $\rho_{\rm ab}$ in various magnetic fields for LBCO with $x=0.08$, 0.10, 0.11 and LNSCO with $x=0.12$. 
For $x=0.10$, 0.11 and 0.12, a jump in $\rho_{\rm ab}$ is observed at $T_{\rm d2} \sim 41$ K, $\sim 51$ K and $\sim 67$ K with decreasing temperature, respectively. 
For $x=0.08$, no jump is observed, suggesting that no structural transition to the TLT phase occurs and that the OMT phase remains at least down to the lowest measured temperature of 1.5 K.~\cite{goka} 

\begin{figure}[tbp]
\begin{center}
\includegraphics[width=1.0\linewidth]{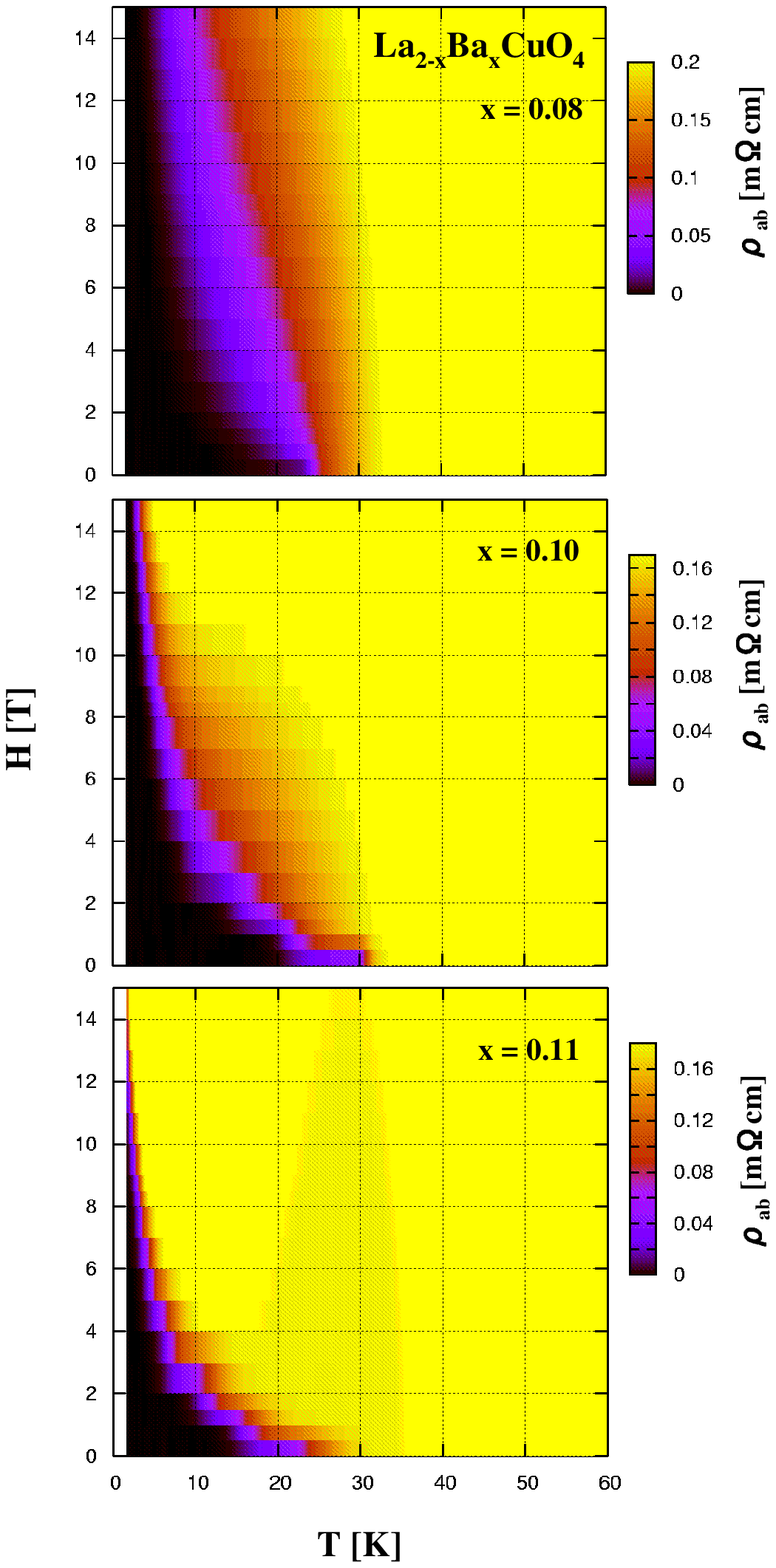}
\end{center}
\caption{(color online) Contour maps of $\rho_{\rm ab}$ in the $H$ vs $T$ plane for La$_{2-x}$Ba$_x$CuO$_4$ with $x=0.08$, 0.10 and 0.11.}  
\label{contour} 
\end{figure}

Focusing the attention on the superconducting transition curve, the broadening is observed with increasing field for $x=0.08$, as usually observed in the underdoped high-$T_{\rm c}$ cuprates. 
For $x=0.11$, on the other hand, it is found that the superconducting transition curve shifts to the low-temperature side in parallel with increasing field.~\cite{lbco-jpcs} 
To be more visible, contour maps of $\rho_{\rm ab}$ in the $H$ vs $T$ plane are shown in Fig. \ref{contour} for LBCO with $x=0.08$, 0.10 and 0.11. 
In each $x$, the region where the color starts to change from that at high temperatures of $\sim$ 60 K with decreasing temperature roughly corresponds to the onset region of the superconducting transition. 
For $x=0.08$, it is found that the relatively sharp transition of $\rho_{\rm ab}$ around $T_{\rm c}$ in zero field is broadened with increasing field. 
For $x=0.11$, on the contrary, the transition of $\rho_{\rm ab}$ around $T_{\rm c}$ remains sharp even in magnetic fields. 
The parallel shift of $\rho_{\rm ab}$ is also observed for $x=0.12$ in LNSCO as shown in Fig. \ref{r-t}. 
These suggest an intimate relation between the parallel shift and the suppression of superconductivity around $p=1/8$ or the formation of the charge-spin stripe order. 

A remarkable feature is for $x=0.10$ that the superconducting transition curve shows the broadening in low fields, while it changes to the parallel shift in high fields above 9 T. 
In Fig. \ref{contour}, it is found that the broad transition of $\rho_{\rm ab}$ around $T_{\rm c}$ below 9 T changes to the sharp one above 9 T for $x=0.10$.
This dramatic change indicates that the application of magnetic field causes a crossover from the usual state of the underdoped high-$T_{\rm c}$ cuprates to the peculiar state around $p=1/8$. 

Another remarkable feature observed for $x=0.10$ is that the normal-state-like behavior of $\rho_{\rm ab}$, characterized by the almost linear $T$-dependence, is observed between $T_{\rm d2}$ and the onset temperature of superconductivity, $T_{\rm c}^{\rm onset}$, of $\sim 15$ K at 9 T and that $\rho_{\rm ab}$ between $T_{\rm d2}$ and $T_{\rm c}^{\rm onset}$ increases with increasing field above 9 T and finally exhibits an insulating behavior for $H \ge 13$ T. 
For $x=0.11$ and 0.12, on the other hand, the increase of $\rho_{\rm ab}$ between $T_{\rm d2}$ and $T_{\rm c}^{\rm onset}$ with increasing field is negligibly small up to 15 T, compared with that for $x=0.10$. 
To be summarized, LBCO with $x=0.08$ is a typical underdoped sample characterized by the broadening of the superconducting transition curve in magnetic fields. 
Both LBCO with $x=0.11$ and LNSCO with $x=0.12$ are peculiar samples around $p=1/8$ characterized by the parallel shift of the superconducting transition curve in magnetic fields. 
LBCO with $x=0.10$ is a rather unique sample that shows the broadening in low fields and the parallel shift in high fields above 9 T and whose $\rho_{\rm ab}$ in the normal state below $T_{\rm d2}$ markedly increases with increasing field up to 15 T. 

%*****************************************************************************************
\section{Discussion}
%*****************************************************************************************
First, we discuss the intimate relation between the superconducting transition curve and the stripe order. 
It is well-known that the broadening of the superconducting transition curve observed for $x=0.08$ is characteristic of the underdoped high-$T_{\rm c}$ cuprates with large superconducting fluctuation.~\cite{kitazawa,suzuki} 
On the other hand, the parallel shift of the superconducting transition curve is observed for $x=0.11$ and 0.12 and for $x=0.10$ above 9 T. 
The parallel shift has also been observed in LNSCO with $x=0.15$ (Ref. \cite{wakimoto}) and LSCO with $x=0.12$ (Ref. \cite{suzuki-san2}) where the static stripe order of charges and/or spins is formed at low temperatures. 
These suggest that both the superconductivity with small superconducting fluctuation and the static stripe order are realized in one sample. 
Here, it is an important issue whether the superconducting region and the static stripe-ordered one coexist microscopically or are separated macroscopically. 
From the neutron scattering measurements in La$_{1.875}$Ba$_{0.125-x}$Sr$_x$CuO$_4$, the static stripe order has been suggested to compete with the superconductivity.~\cite{lbsco} 
Moreover, from the muon-spin-relaxation measurements in La$_{2-x}$Sr$_x$Cu$_{1-y}$Zn$_y$O$_4$ around $x=1/8$, it has been suggested that the superconductivity is destroyed in a region where frequencies of the dynamical stripe fluctuations are lower than $\sim 10^{11}$ Hz.~\cite{musr} 
Therefore, the superconducting region and the static stripe-ordered one are probably separated macroscopically. 

The reason why the superconductivity with small superconducting fluctuation is realized in a sample with the static stripe-ordered region is still an open question. 
A possible origin is that the out-of-plane superconducting coherence length, $\xi_{\rm c}$, might be relatively large under the influence of the correlation of the static stripe order along the c-axis,~\cite{prb,kivelson} leading to the three-dimensional superconductivity with small superconducting fluctuation. 

Next, we discuss the relation between the change of the normal-state behavior of $\rho_{\rm ab}$ in magnetic fields and the field-induced stripe order. 
Considering the results of the elastic neutron scattering measurements~\cite{nature,prb,goka,fujita} mentioned in Sec. \ref{intro}, the negligibly small increase of $\rho_{\rm ab}$ below $T_{\rm d2}$ with increasing field for $x=0.11$ and 0.12 seems to indicate that the stripe order is nearly perfectly stabilized in zero field and is insensitive to the applied field.~\cite{wakimoto} 
As for $x=0.10$, the marked increase of $\rho_{\rm ab}$ below $T_{\rm d2}$ with increasing field reminds us some possible origins. 
The first is the normal-state magnetoresistance. 
However, this is not applicable, because the normal-state magnetoresistance is usually as small as an order of 1 \% at 15 T, as in the case of $x=0.11$ and 0.12. 
The second is the suppression of the superconducting fluctuation by the applied field. 
The positive magnetoresistance appears even at temperatures just below $T_{\rm d2}$ far from $T_{\rm c}^{\rm onset}$ in high fields. 
On the other hand, reports on the Nernst effect in LSCO have suggested that the vortex state survives even at higher temperatures far above $T_{\rm c}$, meaning that the superconducting fluctuation exists even at high temperatures.~\cite{xu,wang} 
Therefore, this possible origin cannot be excluded and further measurements are needed to conclude. 
Nevertheless, it appears that this is not a candidate, because the large positive magnetoresistance is observed only for $x=0.10$ and not for $x=0.11$ nor 0.12. 
The third is enhancement of the localization of holes induced by the applied field. 
The localization behavior of $\rho_{\rm ab}$ becomes marked with increasing field and appears to approach the behavior of LNSCO with $x=0.12$ where the stripe order is perfectly stabilized. 
In the long run, the most probable origin is that the charge-spin stripe order is stabilized by the applied field in $x=0.10$. 
This may be the first experimental evidence, to our knowledge, of the {\it charge} stripe order stabilized in magnetic fields. 
To be more conclusive, the neutron scattering measurements in magnetic fields are under way. 

\begin{figure}[tbp]
\begin{center}
\includegraphics[width=1.0\linewidth]{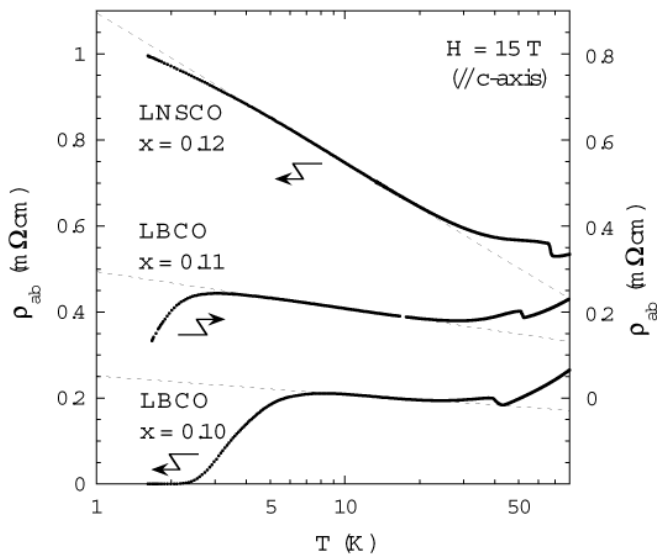}
\end{center}
\caption{Plot of $\rho_{\rm ab}$ vs ln$T$ at 15 T for La$_{2-x}$Ba$_x$CuO$_4$ (LBCO) with $x=0.10$, 0.11 and La$_{1.6-x}$Nd$_{0.4}$Sr$_x$CuO$_4$ (LNSCO) with $x=0.12$.} 
\label{log} 
\end{figure}

Finally, we discuss the temperature dependence of $\rho_{\rm ab}$ below $T_{\rm d2}$. 
So far, it has been clarified in the underdoped LSCO and BSLCO~\cite{ando,boebinger,ono} that $\rho_{\rm ab}$ in the normal state shows the ln(1/$T$) dependence at low temperatures in magnetic fields, suggesting that the ln(1/$T$) dependence is a common feature of the underdoped high-$T_{\rm c}$ cuprates. 
To check this suggestion, $\rho_{\rm ab}$'s of LBCO with $x=0.10$, 0.11 and LNSCO with $x=0.12$ at 15 T are plotted versus ln$T$, as shown in Fig. \ref{log}. 
Below $\sim$ 20 K, it is found that $\rho_{\rm ab}$ in the normal state is proportional to ln(1/$T$) in each $x$, though $\rho_{\rm ab}$ deviates downward from ln(1/$T$) at low temperatures because of the superconducting transition for $x=0.10$ and 0.11. 
The small deviation below 3 K for $x=0.12$ is irrelevant to the superconducting transition, because $\rho_{\rm ab}$ is independent of the field strength for $H\ge$ 11 T, as seen in Fig. \ref{r-t}. 
The slope of the ln(1/$T$) dependence is found to increase with increasing $x$, indicating that the localization of holes  becomes strong with increasing $x$ toward $x=1/8$ at 15 T. 
The ln(1/$T$) dependence is known to be characteristic of the weak localization and the electron-electron interaction in the two-dimensional Anderson-localized state where the in-plane electrical conductivity $\sigma_{\rm ab}$ actually changes in proportion to ln(1/$T$) even in magnetic fields.~\cite{anderson} 
For $x=0.10$, 0.11 and 0.12, however, $\sigma_{\rm ab}$ does not show the ln(1/$T$) dependence below $T_{\rm d2}$, as in the case of the underdoped LSCO.~\cite{ando,boebinger} 
Although the true origin of the ln(1/$T$) dependence of $\rho_{\rm ab}$ is not clear, the localization of holes with the ln(1/$T$) dependence observed widely in the underdoped high-$T_{\rm c}$ cuprates is robust even in the stripe-ordered state of LBCO and LNSCO.~\cite{noda} 

%*****************************************************************************************
\section{Summary}
%*****************************************************************************************
It has been found that the superconducting transition curve shows the parallel shift by the application of magnetic field in LBCO with $x=0.11$ and LNSCO with $x=0.12$ where the charge-spin stripe order is formed at low temperatures. 
These suggest that both the superconductivity with small superconducting fluctuation and the static stripe order are realized in one sample for $x=0.11$ and 0.12. 
For LBCO with $x=0.10$, the broadening in low fields changes to the parallel shift in high fields above 9 T. 
Moreover, $\rho_{\rm ab}$ in the normal state below $T_{\rm d2}$ increases with increasing field up to 15 T. 
It is possible that these pronounced features of $x=0.10$ are understood in terms of the field-induced stabilization of the {\it charge} stripe order. 
The $\rho_{\rm ab}$ in the normal state at low temperatures has been found to be proportional to ln(1/$T$) for $x=0.10$, 0.11 and 0.12, suggesting the localization of holes with the ln(1/$T$) dependence of $\rho_{\rm ab}$ which is robust even in the stripe-ordered state of LBCO and LNSCO.

%*****************************************************************************************
\section*{Acknowledgments}
%*****************************************************************************************
The high magnetic field experiments were partly supported by the High Field Laboratory for Superconducting Materials (HFLSM), Institute for Materials Research, Tohoku University. 
This work was supported by a Grant-in-Aid for Scientific Research from the Ministry of Education, Science, Sports, Culture and Technology, Japan.

\end{document}